\documentclass[cits]{PoS}
\usepackage{graphicx}

\title{Magnetic fields in nearby galaxies:\\ prospects with future radio telescopes}

\ShortTitle{Magnetic fields in nearby galaxies}

\author{\speaker{Rainer Beck}\\
        Max-Planck-Institut f\"ur Radioastronomie, Auf dem H\"ugel 69, 53121 Bonn, Germany\\
        E-mail: \email{rbeck@mpifr-bonn.mpg.de}}

\abstract{The origin of magnetic fields in the Universe is an open
problem in astrophysics and fundamental physics. Our present-day
knowledge is limited to regions of strong magnetic fields and to
star-forming disks of galaxies. Cosmic-ray electrons emitting at
high (GHz) radio frequencies can propagate only a few kpc from their
places of origin due to their lifetime limited by energy losses.
Low-energy electrons emitting at low frequencies suffer less from
energy losses and can propagate further into the intergalactic
medium. The prospects are threefold: Firstly, LOFAR will map the
structure of weak magnetic fields in the outer regions and halos of
galaxies and in the Milky Way. Polarized emission is an excellent
tracer of past interactions with other galaxies and with the
interstellar medium. Secondly, high-resolution polarization
observations are needed at high frequencies with the EVLA and SKA to
trace the structure of magnetic fields in the disks and central
regions of galaxies in unprecedented detail. The SKA can also detect
polarized emission from distant, unresolved galaxies. Thirdly,
Faraday rotation measures (RM) are signatures of regular magnetic
fields generated by the dynamo mechanism. All-sky surveys of Faraday
rotation measures (RM) towards polarized background sources will be
used to model the structure and strength of the regular magnetic
fields in the Milky Way, the interstellar medium of galaxies and the
intergalactic medium. The novel method of {\em RM Synthesis},
applied to spectro-polarimetric data cubes, is able to separate RM
components from different distances and may allow 3-D {\em Faraday
tomography}. This will open a new era in the observation of cosmic
magnetic fields. ``Key Science'' Projects on cosmic magnetism are
organized for the Low Frequency Array (LOFAR), the planned Square
Kilometre Array (SKA) and the Australian SKA Pathfinder telescope
(ASKAP). }

\FullConference{Panoramic Radio Astronomy: Wide-field 1-2 GHz research
on galaxy evolution - PRA2009\\
         June 02 - 05 2009\\
         Groningen, the Netherlands}

\begin{document}

\section{Origin of galactic magnetic fields}

The origin of the first magnetic fields in the Universe is still a
mystery \cite{widrow02}. A large-scale primordial field is hard to
maintain in a young galaxy because the galaxy rotates
differentially, so that field lines get strongly wound up during
galaxy evolution, while observations show significant pitch angles.
Turbulent ``seed'' fields in young galaxies can originate from the
Weibel instability in shocks during the cosmological structure
formation \cite{lazar09} or can be injected by the first stars or
jets generated by the first black holes \cite{rees05}. In any case a
mechanism to sustain and organize the magnetic field is required.

The most promising mechanism is the dynamo \cite{beck96,brand05}
which transfers mechanical into magnetic energy. In young galaxies
without ordered rotation a small-scale dynamo \cite{brand05}
probably amplified the seed fields from the protogalactic phase to
the energy density level of turbulence within less than $10^9$~yr
\cite{arshakian09} (see also this volume). To explain the generation
of large-scale fields, the {\em mean-field $\alpha\Omega$--dynamo}
has been developed as a simplified model. It needs turbulence,
differential rotation and helical gas flows ($\alpha$ effect),
generated by supernova explosions \cite{ferriere00,gressel08} or
cosmic-ray driven Parker loops \cite{hanasz04,moss99,parker92}. The
mean-field $\alpha\Omega$--dynamo in galaxy disks predicts that
within a few $10^9$~yr large-scale regular fields are excited from
the seed fields, described as ``modes'' with different azimuthal
symmetry in the disk and vertical symmetry perpendicular to the disk
plane \cite{arshakian09,beck96}.

The mean-field dynamo generates large-scale helicity with a non-zero
mean in each hemisphere. As total helicity is a conserved quantity,
the dynamo is quenched by the small-scale fields with opposite
helicity unless these are removed from the system \cite{shukurov06}.
It seems that {\em outflows}\ are essential for an effective
mean-field dynamo.

Classical dynamo models are strongly simplified because small and
large scales are artificially separated and the back-reaction of the
field onto the gas flow is not included. Dynamical MHD models of
rotating galaxy disks where dynamo action at all scales emerges
automatically are just becoming feasible with present-day computers
\cite{hanasz09}.

\begin{figure*}[t]
\vspace*{2mm}
\begin{minipage}[t]{7cm}
\begin{center}
\includegraphics[bb = 32 99 537 702,width=6cm,clip=]{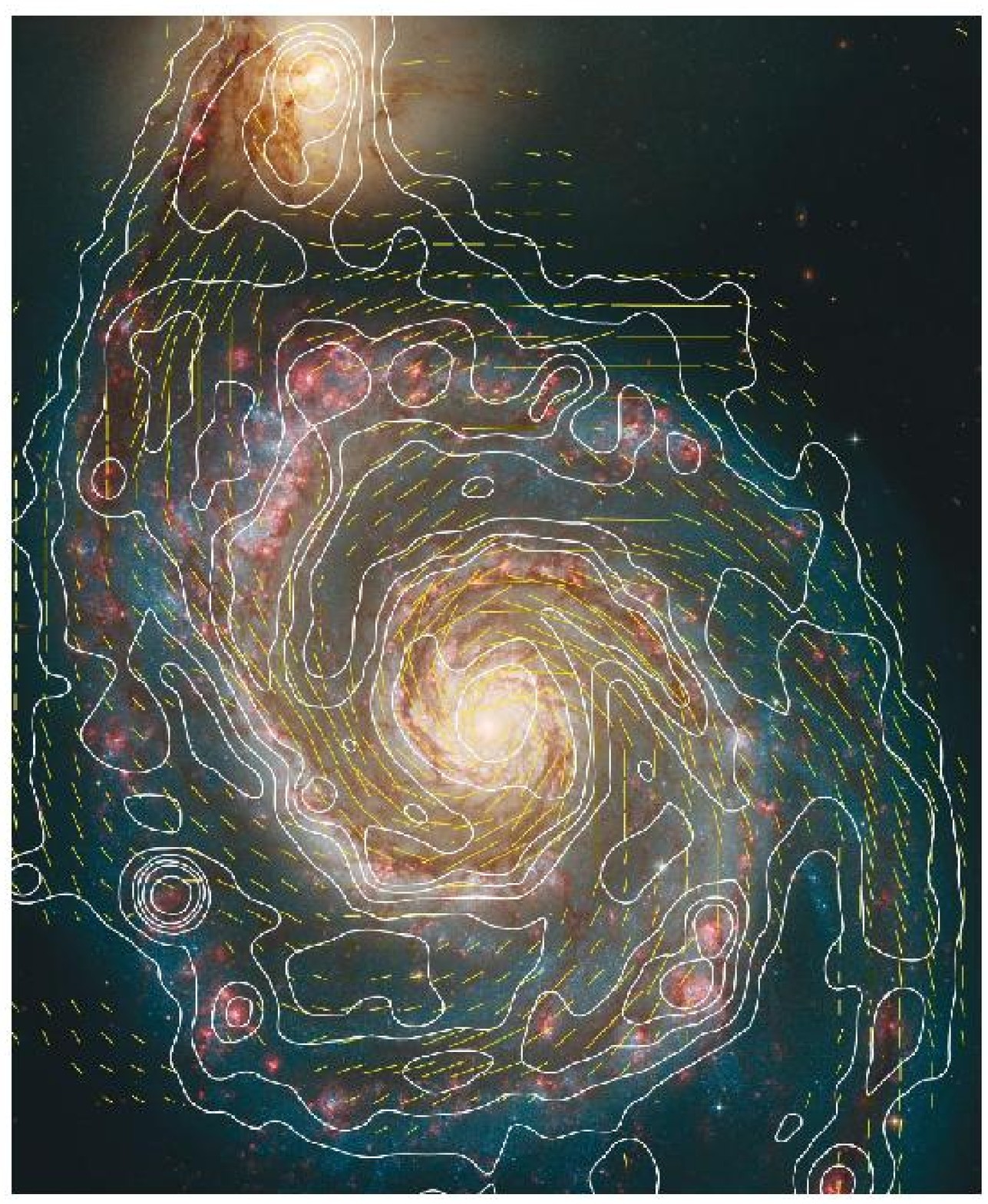}
\caption{Total radio emission (contours) and $B$--vectors of M~51
observed at 6~cm wavelength with the VLA and Effelsberg telescopes
and smoothed to 15''\ resolution \cite{fletcher09}, overlaid onto an
optical image from the HST (Copyright: MPIfR Bonn and \emph{Hubble
Heritage Team}. Graphics: \emph{Sterne und Weltraum}).}
\label{fig:m51}
\end{center}
\end{minipage}\hfill
\begin{minipage}[t]{7cm}
\begin{center}
\includegraphics[bb = 108 150 464 643,width=6cm,clip=]{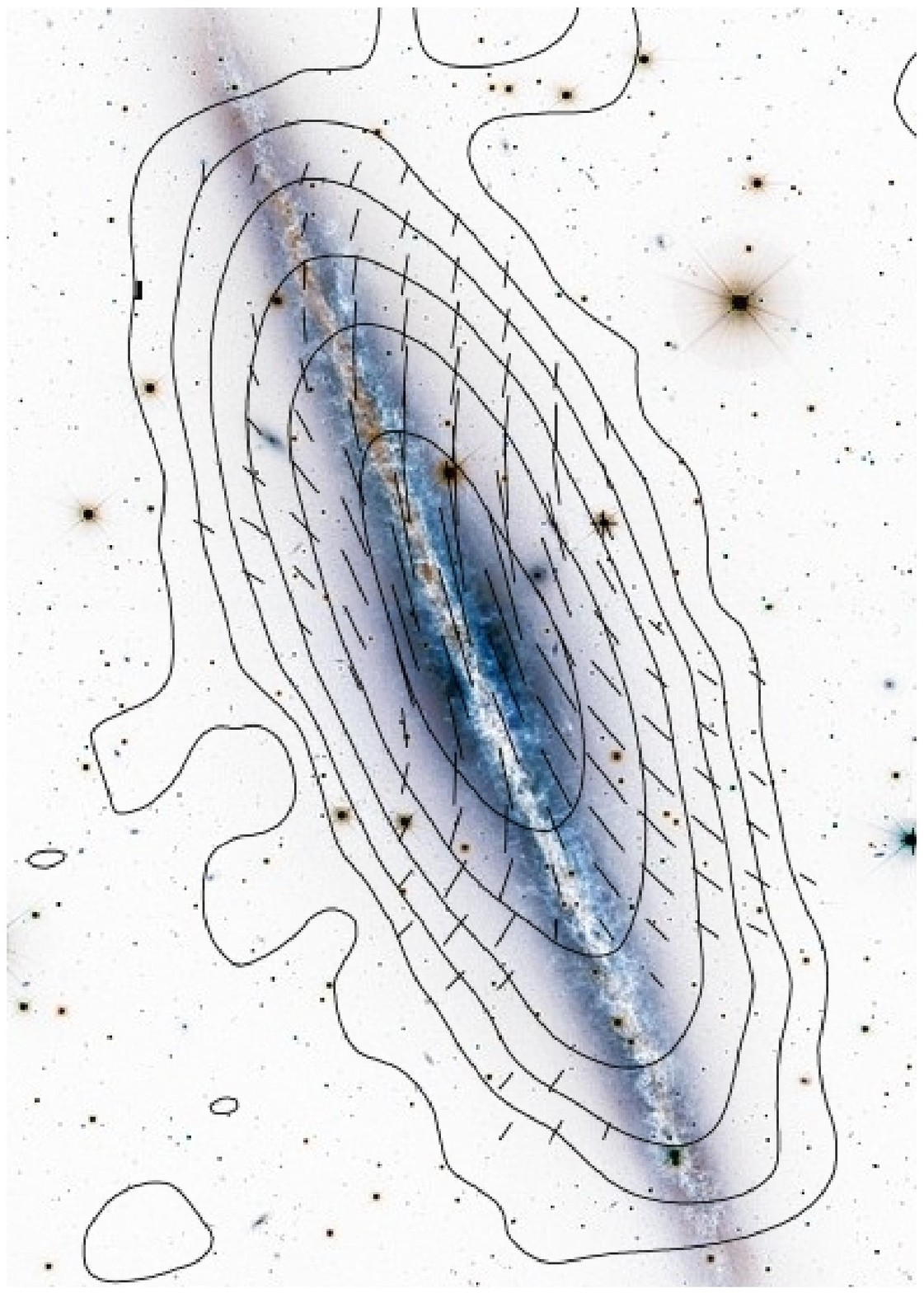}
\caption{Total radio emission (84''\ resolution) and $B$--vectors of
the edge-on spiral galaxy NGC~891, observed at 3.6~cm wavelength
with the Effelsberg 100m telescope \cite{krause08}. The background
optical image is from the CFHT (Copyright: MPIfR Bonn and
CFHT/Coelum).} \label{fig:n891}
\end{center}
\end{minipage}
\end{figure*}

\section{Present-day knowledge about galactic magnetic fields}

The typical average ``equipartition'' strength of the total magnetic
field in spiral galaxies is about $10~\mu$G, assuming energy
equipartition between cosmic rays and magnetic fields. Radio-faint,
gas-poor galaxies like M~31 have weaker total magnetic fields (about
$5~\mu$G on average, higher in star-forming regions), while gas-rich
spiral galaxies with high star-formation rates, like M~51
(Fig.~\ref{fig:m51}) and NGC~6946, have total field strengths of
15~$\mu$G on average and 20--30~$\mu$G in their spiral arms. The
mean energy density of the equipartition magnetic field (and hence
that of the cosmic rays) is $\simeq10^{-11}$~erg~cm$^{-3}$ in
NGC~6946 and $\simeq10^{-12}$~erg~cm$^{-3}$ in M~33
\cite{beck07,taba08}, about 10 times larger than that of the ionized
gas, but similar to that of the turbulent gas motions across the
whole star-forming disk.

The strongest total fields of 50--100~$\mu$G were found so far in
starburst galaxies, like M~82 and the ``Antennae'' NGC~4038/9
\cite{chyzy04}, and in nuclear starburst regions, like in the
centers of NGC~1097 and other barred galaxies \cite{beck+05}. In
starburst galaxies the equipartition field strength is probably
underestimated due to strong energy losses of the cosmic rays
\cite{thompson06} which was recently confirmed by Zeeman
measurements of OH maser lines \cite{robishaw08}.

The degree of radio polarization within the spiral arms is only a
few \%; hence the field in the spiral arms is mostly tangled or
randomly oriented within the telescope beam, the width of which
corresponds to scales between a few 100~pc and a few kpc. Turbulent
fields in spiral arms are probably generated by turbulent gas
motions related to star formation activity.

The ordered (regular and/or anisotropic) fields traced by the
polarized synchrotron emission are generally strongest
(10--15~$\mu$G) in the regions {\em between}\ the optical spiral
arms and oriented parallel to the adjacent spiral arms, in some
galaxies forming {\em magnetic arms}. These are probably generated
by the mean-field dynamo \cite{beck96}. However, in galaxies with
strong density waves like M~51 (Fig.~\ref{fig:m51}) the turbulent
field is compressed at the inner edge of the spiral arms and becomes
anisotropic, while the regular dynamo field is weak
\cite{fletcher09}. The ordered magnetic field forms spiral patterns
in almost every galaxy \cite{beck05} and also in the central regions
of galaxies and circum-nuclear gas rings \cite{beck+05}. Spiral
fields can be generated by compression at the inner edge of spiral
arms, by shear in interarm regions, or by dynamo action
\cite{beck96}.

Large-scale patterns of Faraday rotation measures (RM) are
signatures of regular dynamo fields and can be inferred from
polarized emission of the galaxy disks \cite{krause90} or from RM
data of polarized background sources \cite{stepanov08}. The
Andromeda galaxy M~31 hosts a dominating axisymmetric disk field
\cite{fletcher04}, as predicted by dynamo models.
However, in many observed galaxy disks
no clear patterns of Faraday rotation were found. Either several
dynamo modes are superimposed and cannot be distinguished with the
limited sensitivity and resolution of present-day telescopes, or the
timescale for the generation of large-scale modes is longer than the
galaxy's lifetime, or the regular field was distorted by interaction
or a major merger \cite{arshakian09}.

Nearby galaxies seen ``edge-on'' generally show a disk-parallel
field near the disk plane. High-sensitivity observations of edge-on
galaxies like NGC~891 (Fig.~\ref{fig:n891}) and NGC~253
\cite{heesen09} revealed vertical field components in the halo
forming an X-shaped pattern. The field is probably transported from
the disk into the halo by a gas outflow emerging from the disk.

\section{Limitations of observations with present-day
radio telescopes}

If the magnetic field is dynamically important, dynamo MHD models
are needed that include the back-reaction onto the gas flow.
Detailed comparisons with observations of the field structure and
the distributions and velocities of the various gas components need
spatial resolutions of better than 100~pc. Present-day radio
telescopes do not provide sufficient sensitivity at such resolutions
because the observed signal from extended sources decreases with the
beam area.

The observation of synchrotron emission from galaxies only reveals
magnetic fields illuminated by cosmic-ray electrons (CRE)
accelerated by supernova shock fronts in regions of strong star
formation. As the diffusion length of CRE is limited by the Alfv\'en
speed and the lifetime due to synchrotron and inverse Compton
losses, radio images at centimeter wavelengths (synchrotron +
thermal) are similar to images of star-forming regions as observed
in the far-infrared. Radio emission from the outer disks and halos
of galaxies is very weak due to the lack of emitting electrons.
Hence, the strength and extent of magnetic fields into the
intergalactic space is unknown.

Asymmetric distributions of the polarized emission from galaxies in
the Virgo cluster show that the outer magnetic fields have been
strongly compressed \cite{vollmer07}. Polarized emission is an
excellent tracer of past interactions between galaxies or with the
intergalactic medium, but the observation is still difficult.

Faraday rotation is a signature of regular fields as generated by
dynamo action, but the existing data are still scarce and often do
not reveal clear patterns. Higher angular resolution is needed to
distinguish between a spectrum of superimposed dynamo modes and
anisotropic (sheared) fields. Most of the existing RM data also
suffer from low spectral resolution causing depolarization by
different Faraday rotation components within the beam or along the
line of sight. {\em RM Synthesis}\ is capable to reveal such
multiple RM components but needs a large number of channels over a
wide frequency range \cite{brentjens05,heald09}.

RM data from pulsars in the Milky Way indicate several field
reversals at various Galactic radii \cite{noutsos09}. However, no
model of a large-scale field structure is statistically safe
\cite{men08}. Similar to external galaxies, the Milky Way's regular
field probably has a complex structure that cannot be resolved with
the existing number of pulsar RM.

\section{Prospects}

Future radio telescopes will widen the range of observable magnetic
phenomena. High-resolution, deep observations at high frequencies,
where Faraday effects are small, require a major increase in
sensitivity for continuum observations which will be achieved by the
Extended Very Large Array (EVLA) and the planned Square Kilometre
Array (SKA). We wish to resolve the detailed structure of the ISM
and halo fields and distinguish sheared loops from regular fields
with dynamo-type patterns \cite{stepanov08}. The SKA will also allow
to measure the Zeeman effect in much weaker magnetic fields in the
Milky Way and in nearby galaxies.

Forthcoming low-frequency radio telescopes like the Low Frequency
Array (LOFAR), Murchison Widefield Array (MWA), Long Wavelength
Array (LWA) and the low-frequency SKA will be suitable instruments
to search for extended synchrotron radiation at the lowest possible
levels in outer galaxy disks and clusters and the transition to
intergalactic space \cite{beck08}. The detection of radio emission
from the intergalactic medium would allow us to probe the existence
of magnetic fields in such rarified regions, measure their
intensity, and investigate their origin and their relation to the
structure formation in the early Universe.

Nearby galaxies seen edge-on generally show a disk-parallel field
near the disk plane, so that polarized emission can also be detected
from unresolved galaxies \cite{stil09} (see also this volume). With
the SKA ordered fields in young galaxies can be searched for.

Radio spectro-polarimetric observations in many narrow frequency
channels allows application of {\em RM Synthesis}\
\cite{brentjens05,heald09}. Faraday depolarization occuring in wide
frequency bands is reduced and features at different distances along
the line of sight can be separated. If the medium has a relatively
simple structure, for example a few emitting regions and Faraday
screens, {\em Faraday tomography}\ will become possible. This method
is going to revolutionize radio polarization observations.

A reliable model for the global structure of the magnetic field of
the Milky Way and nearby galaxies needs a much higher number of
pulsar and extragalactic RM, hence larger sensitivity and/or higher
survey speed. The POSSUM-wide survey at 1.4~GHz with the planned
Australia SKA Pathfinder (ASKAP) telescope will measure about 80 RM
from polarized extragalactic sources per square degree, the
POSSUM-deep survey about 10 times more. With 12h integration at
1.4~GHz, the SKA will be able to detect 1~$\mu$Jy sources and
measure about 8000 RM per square degree. The SKA ``Magnetism'' Key
Science Project plans to observe an all-sky RM grid with 1h
integration per field which should contain about $10^4$ RM values
from pulsars with a mean spacing of $\simeq30'$ and about $10^8$ RM
from compact polarized extragalactic sources at a mean spacing of
just $\simeq1.5'$ \cite{gaensler04}. This survey will be used to
model the structure and strength of the magnetic fields in the
foreground of the Milky Way, in intervening galaxies, and in the
intergalactic medium \cite{beck04}. 10 RM values in the solid angle
area of a foreground object are already sufficient to recognize a
simple large-scale field structure, while more than 1000 values are
required for a detailed field reconstruction \cite{stepanov08}. As
the accuracy depends on the polarized flux of the background source,
the distance range of this method is much larger than by direct
imaging of the polarized emission from the intervening galaxy.
Looking back into time, the future telescopes can shed light on the
origin and evolution of cosmic magnetic fields.

\end{document}